\documentclass{article}

\usepackage{arxiv}
\usepackage{makecell}
\usepackage[utf8]{inputenc} % allow utf-8 input
\usepackage[T1]{fontenc}    % use 8-bit T1 fonts
\usepackage{hyperref}       % hyperlinks
\usepackage{color}
\usepackage[figuresright]{rotating}
\usepackage{url}            % simple URL typesetting
\usepackage{booktabs}       % professional-quality tables
\usepackage{amsfonts}       % blackboard math symbols
\usepackage{nicefrac}       % compact symbols for 1/2, etc.
\usepackage{microtype}      % microtypography
\usepackage{lipsum,amsmath}
\usepackage{graphicx}
\usepackage{natbib}
\usepackage{algpseudocode,algorithm,algorithmicx}
\newcommand{\indep}{\perp \!\!\! \perp}

\setcitestyle{authoryear,open={(},close={)}}

\usepackage{setspace}
\title{SPORTSCausal: Spill-Over Time Series Causal Inference}

\author{
  Carol Liu \\
  Department of Marketing\\
  University of Wisconsin-Madison\\
  Madison, WI, USA\\
  \texttt{liu747@wisc.edu} \\
}

\begin{document}
\maketitle

\begin{abstract}

Randomized controlled trials (RCTs) have long been the gold standard for causal inference across various fields, including business analysis, economic studies, sociology, clinical research, and network learning. The primary advantage of RCTs over observational studies lies in their ability to significantly reduce noise from individual variance. However, RCTs depend on strong assumptions, such as group independence, time independence, and group randomness, which are not always feasible in real-world applications. Traditional inferential methods, including analysis of covariance (ANCOVA), often fail when these assumptions do not hold. In this paper, we propose a novel approach named \textbf{Sp}ill\textbf{o}ve\textbf{r} \textbf{T}ime \textbf{S}eries \textbf{Causal} (\verb+SPORTSCausal+), which enables the estimation of treatment effects without relying on these stringent assumptions. We demonstrate the practical applicability of \verb+SPORTSCausal+ through a real-world budget-control experiment. In this experiment, data was collected from both a 5\% live experiment and a 50\% live experiment using the same treatment. Due to the spillover effect, the vanilla estimation of the treatment effect was not robust across different treatment sizes, whereas \verb+SPORTSCausal+ provided a robust estimation.

\end{abstract}

\keywords{\\Randomized Control Trial \and Spill-over Effect \and Causal Inference}

\section{Introduction and related work}

Randomized controlled trial(RCT) has been playing a central role in causal inference. The easiest design of RCT is single-treatment complete randomized design (CRD), which is popular in business analysis and also clinical investigation. More modern design, such as randomized complete block design (RCBD), factorial design(\citep{mukerjee2007modern}), latin-square design(\citep{bradley1958complete}), suduko square design(\citep{sarkar2015sudoku}) etc are also constructed for multi-factor complicated experiments. 

Mixed effect model (as R package \verb+lme4+) was designed as a comprehensive tool for all of those experimental designs (\citep{bates2007lme4}).It is simple as analysis of covariance (ANCOVA) when there is no confounding blocks or unit-level random covariates. Take the single-treatment CRD as an example, the averaged treatment effect and its corresponding statistical inference could be evaluated by regress the time $T$ outcome with time 0 outcome and the treatment indicator ($y_{it} \sim y_{i0} + d_i$ where $d_i$ is the treatment indicator). $y_{it}$ and $y_{0t}$ are always substituted by the average over time "after" and "before" the treatment. 

However, similar as \cite{rubin1986comment} described, most experimental designs require the following three assumptions for the traditional methodologies to be valid:
\begin{itemize}
    \item Group randomness: the random design when grouping samples into treatment groups and control groups.
    \item Time independence: the treatment effect is not time dependent.
    \item Independence between treatment groups: during the period of experiment, there is no interaction among groups.
\end{itemize}

These assumptions simplify statistical inference but are often not valid in real-world settings. Statisticians continue to explore various modeling strategies to relax these assumptions and create more flexible models.

Grouping randomness is a key assumption in experimental design, yet it is not always achievable due to constraints in the experimental environment and sample size. For instance, as discussed in \citep{zheng2024bootstrap}, in business analysis, a common approach is the tail-number experiment, where subjects with odd-numbered sample IDs are placed in the experimental group, while those with even-numbered IDs are placed in the control group. This design is random only if no other experiments are concurrently using similar ID tail-number information. Large sample theory ensures that when grouping randomness is present, all features are balanced between groups, reducing noise from unit-level variability and confounding covariates. Without grouping randomness, the experiment offers little advantage over observational studies, potentially leading to differences between the experimental and control groups even before treatment begins.

To address this issue, various methods have been developed, such as Difference-in-Differences (DID, \citep{luke1998interracial}), Propensity Score Matching (PSM, \citep{rosenbaum1985constructing}), and Synthetic Control (\citep{abadie2010synthetic}), all of which aim to match experimental and control subjects on pre-treatment characteristics. Recent advancements, such as \citep{zheng2024bootstrap} Bootstrap Matching, have been made to enhance the robustness of PSM in experimental settings.

Time independence is another critical assumption in experimental design, as it simplifies the inference process by avoiding the complexities of longitudinal time series data. Researchers often prefer longer pre- and post-treatment periods to strengthen their assumptions, although they typically average over time rather than fully utilizing the longitudinal information. However, approaches like those by \cite{robins2000marginal} in epidemiology and the \verb+causalimpact+ R package developed by Google for Bayesian longitudinal causal inference \citep{brodersen2015inferring} have started to integrate time series data into causal analysis.

The assumption of independence between treatment groups, encapsulated in the "Stable Unit Treatment Value Assumption" (SUTVA, \citep{rubin1986comment}), posits that the treatment should only affect the experimental group and not the control group, with no reciprocal influence between the two. While this assumption generally holds in many study areas, there are situations where its violation leads to significant issues in causal inference. For instance, in business analysis, particularly in advertising strategy, if group A is subjected to a traditional strategy and group B to a new algorithm, the competition dynamics between the two groups can cause spillover effects. These effects occur when changes in group B indirectly influence group A, complicating the interpretation of the treatment effect. The observed difference between the groups is then a combination of both the direct (treatment) and indirect (spillover) effects. \cite{aronow2020spillover} explored methods to estimate these spillover effects, and R packages like \verb+inferference+ and \verb+interferenceCI+ have been developed to address similar challenges. However, these methods often require sophisticated pseudo-experimental designs to isolate the behavior of the control group during treatment, without the interference of spillover effects.

In this paper, we introduce a novel statistical method, \verb+SPORTSCausal+, designed to perform causal inference in scenarios where group randomness is absent, but spillover effects are present. We demonstrate the application of \verb+SPORTSCausal+ using a real-world budget-control experiment, highlighting its effectiveness in handling complex causal relationships.

\section{Methodology and Algorithm}

\subsection{Notations and key assumptions}
Let's formalize the question, taking a single treatment RCT as an example. More complicated design could be easily generated here to capture the spillover effect. Denote $i = 1,2,\cdots,m, m+1, \cdots, m+n$ be the number of subjects where the first $m$ subjects are randomly chosen to fit in the base group and the rest $n$ subjects in experiment group. Suppose $t = 1,2,\cdots, T_0$ be the time period before the treatment and $t = T_0 + 1, T_0+2, \cdots, T_0 + T$ be the time period after the treatment. Therefore, for the outcome, we have a matrix $Y = \left(y_{it}\right)_{(m+n)*(T_0 + T)}$ where each row is a subject and each column is a time point and $y_{it}$ is for example, the cost/revenue/ROI for subject $i$ at time point $t$. For simplicity, let's denote $D_i$ be the indicator for treatment effect. In other words, $D_i = 0, i\in \{1,2,\cdots,m\}$ and $D_i = 1, i \in \{m+1, \cdots, m+n\}$. We could also have a feature matrix $X = (x_{ij})_{(m+n)*k}$. Supposing each subject has $k$ features (such as age, gender, etc), each row is again a feature vector for subject $i = 1,2,\cdots, m+n$ and $j = 1,2,\cdots, k$. Finally the unknown treatment effect to be $d$.

Let's define previous several assumptions:

\begin{enumerate}
    \item (\textbf{$C_1$ Group randomness}): \\
    $\forall i \neq j, \mathbb{P}(D_i = 1) = \mathbb{P}(D_j = 1)$.
    \item (\textbf{$C_2$ Time independence}): \\
    $d \indep T$.
    \item (\textbf{$C_3$ Independence between treatment groups}): \\
    $\forall t\geq T_0 + 1$, $\forall i \in \{1,2,\cdots, m\}$ and $\forall j \in \{m+1,\cdots, m+n\}$, $\mathbb{P}(y_{it}) = \mathbb{P}(y_{it}|y_{jt})$
\end{enumerate}

Let's here make a few remarks:

\begin{enumerate}
    \item Under assumption $C_1$, large sample theory guarantees that the conditional distribution is same between groups $\mathbb{P}(X_{ij}|D_i = 1) = \mathbb{P}(X_{ij}|D_i = 0)$. It also holds for outcome values when $t\leq T_0$, i.e., $\mathbb{P}(y_{it}|D_i = 0) = \mathbb{P}(y_{it}| D_i = 1)$. On the other hand, the violation of assumption $C_1$ fundamentally weaken the benefit of experimental design compared to observational study. 
    \item Under assumption $C_2$, for any $i$, $y_{it}$ could be regarded as a realization independently and identically distributed from a random variable. Therefore, for a relatively long time period, strong law of large number guarantees that $\frac{1}{|t|}\sum_{t}y_{it}$ converge to the true expectation. Note that the summation over $t$ should be divided into two periods, before and after the treatment starts ($T_0$).
    \item Under assumption $C_3$, there is no spillover (indirect) effect between groups, caused by the treatment. Therefore, the difference in behavior between groups can be evaluated as the treatment effect $d$.
\end{enumerate}

The most difficult assumption to be understood is assumption $C_3$. The following figure, Figure~\ref{C3} shows a graphical representation when assumption $C_3$ is satisfied or violated.

\begin{figure}[H] % picture
    \centering
    \includegraphics[width=0.7\columnwidth]{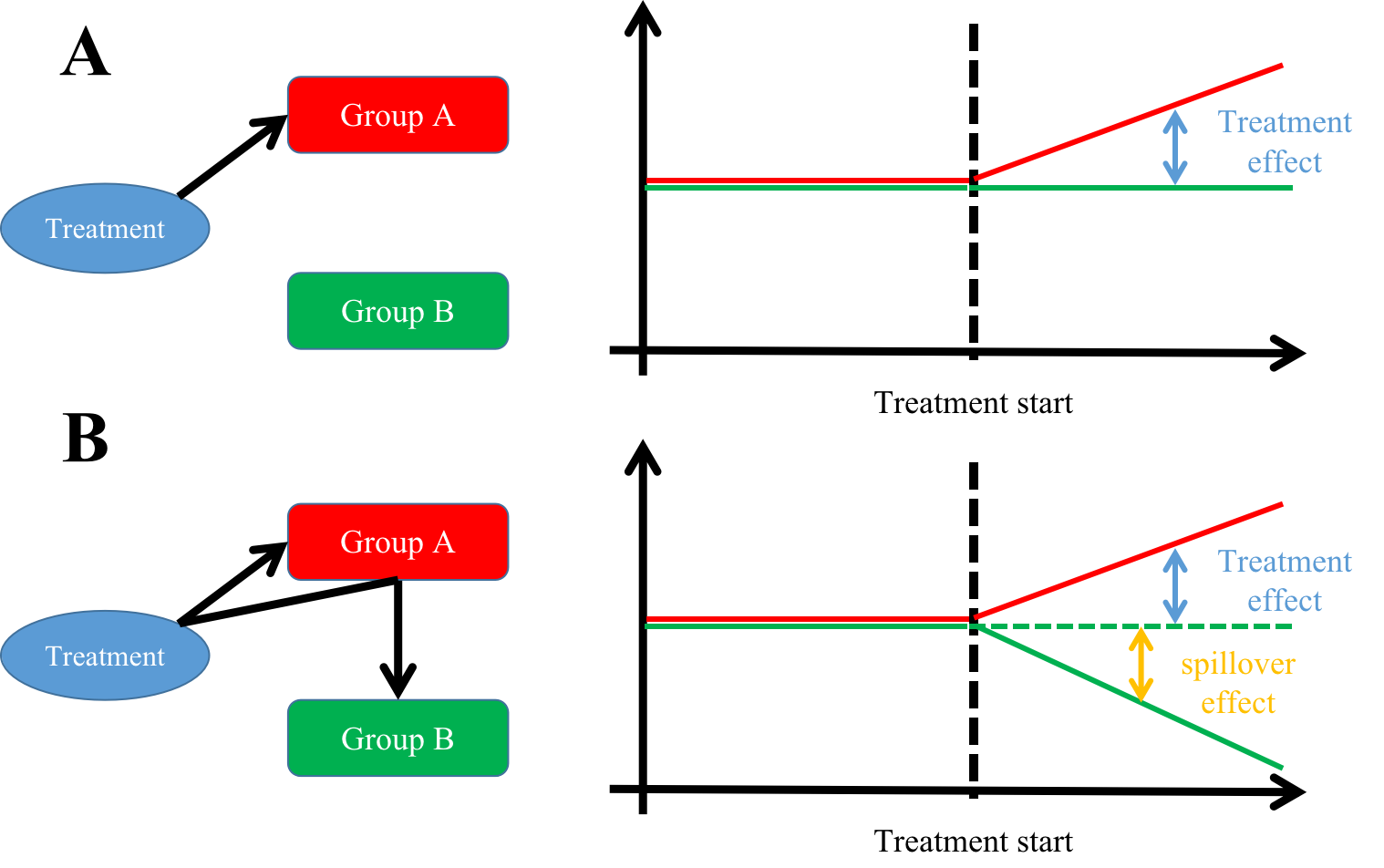}
    \caption{\textbf{When assumption $C_3$ is violated?} Panel A shows how data looks like when $C_3$ is satisfied where panel $B$ shows when it is violated. Left panel shows how treatment influences group A and group B and right panel shows the potential data behavior under different assumptions.}
    \label{C3}
\end{figure}

Several clarifications related to assumption $C_3$:

\begin{itemize}
    \item This is only a simple demonstration showing one potential outcome of treatment (linear in time). However, in reality it might be much complicated.
    \item The influence of treatment into group B is not direct. As shown in left bottom of this figure, the influence of treatment to group B is through group A, as an indirect effect.
    \item If $C_3$ is not satisfied, we can only observe the summation of treatment effect. However we are indeed more interested in the treatment effect. The identifiability of treatment effect from their summation is of importance.
\end{itemize}

Those three assumptions would often be satisfied and people is always utilizing those three assumptions to make easier statistical inferences. We would review different algorithms when the previous assumptions is satisfied or violated. However, before that, we would like to relax assumption $C_2$ and $C_3$ a little bit, that would eventually benefit the algorithm development presented in this paper.

\begin{enumerate}
    \item (\textbf{$C_4$ Constant time series structure}): this is a relaxed version of assumption $C_2$. Under this assumption, we assume that for control group ($\forall i \leq m$) and $\forall t$, $$\mathbb{P}(Y_{it}|Y_{it-1}, Y_{it-2}, \cdots, Y_{it-c}) = \mathbb{P}(Y_{it+1}|Y_{it}, Y_{it-1}, \cdots, Y_{it-c+1})$$ 
    where the probability model could be any time series model and the $c$ presenting here the degree of lag. Note that $C_4$ only holds for control group, since the treatment would definitely be caused by the treatment.
    \item (\textbf{$C_5$ Acyclic interaction between groups during treatment period}): this is a related version of assumption $C_3$. Under assumption $C_5$, we could allow the potential indirect effect of treatment for control group, through the treatment group. However, we do not allow the effect of control group to treatment group. It is equivalent to say, the only influence to treatment group is the treatment itself, rather than the effect from group B. We call this acyclic since if we alow the effect from group A to group B and group B to group A, we have a treatment circle. The circle of treatment would make any estimation unidentifiable.
\end{enumerate}

\begin{figure}[H] % picture
    \centering
    \includegraphics[width=0.3\columnwidth]{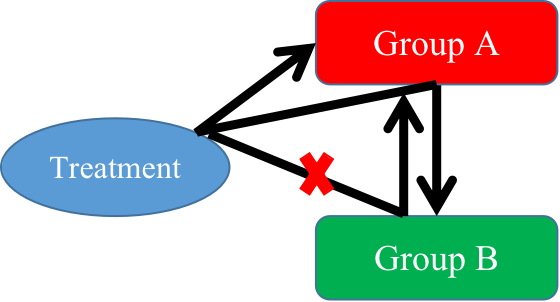}
    \caption{\textbf{What is allowed and not allowed under the assumption $C_5$}}
    \label{C5}
\end{figure}

\subsection{Algorithms under different assumptions}

Statisticians are developing different algorithms under variety of assumptions. Here we only describe a few of them. Our target is to descirbe algorithm only requiring $C_4,C_5$.

\subsubsection{ANCOVA: under $C_1,C_2,C_3$}

\begin{algorithm}[H]
\label{ANCOVA}
  \caption{ANCOVA: under assumption $C_1,C_2,C_3$}
  \hspace*{0.02in} 
  {\bf Input:}  Outcome matrix $Y$, feature matrix $X$, treatment indicating vector $D$\\
  {\bf Output:} Effect size of treatment $d$ and its significance level $p$
  \begin{algorithmic}[1]
  \State Average over time for two groups, before and after the treatment. $$y_{i,after} = \frac{1}{T} \sum_{t = T_0+1}^{T_0+T} y_{it} $$ and similarly, $$y_{i,before} = \frac{1}{T_0} \sum_{t = 1}^{T_0} y_{it}$$
  \State Fit the mixed effect model: $$y_{i,after} \sim y_{i, before} + D_i + X_i + \epsilon_i$$ where different features of $X_i$ can be regarded as either fixed effect or random effect. 
  \State Report estimated coefficient of $D_i$ and the corresponding p-value $p$.
  \end{algorithmic}
\end{algorithm}

This is the easiest algorithm for experimental design. R package \verb+lme4+ can be used to implement this algorithm. Without the subject specific covariates, we can also fit the model using \verb+lm+ or \verb+anova+.

\subsubsection{Bootstrap matching: under $C_2, C_3$}

When $C_1$ is violated, we pretty much have an observational study rather than experimental design. Many statisticians attempt to solve this problem. We describe a recent proposed algorithm \citep{zheng2024bootstrap} name bootstrap matching, which is a robust version of PSM borrowing idea from Bootstrap and PSM.

\begin{algorithm}[H]
\label{Bootstrap matching}
  \caption{Bootstrap matching: under assumption $C_2,C_3$}
  \hspace*{0.02in} 
  {\bf Input:}  Outcome matrix $Y$, feature matrix $X$, treatment indicating vector $D$\\
  {\bf Output:} Effect size of treatment $d$ and its significance level $p$
  \begin{algorithmic}[1]
  \State Repeat the following several steps:
  \State Bootstrap sample from $Y$, denote as $Y^{b}$.
  \State Use bootstrap sample $Y^{b}$ to run propensity score matching based on no matter classifier. Denote the post-matching sample as $\bar Y^{b}$.
  \State Use $\bar Y^{b}$ to run ANCOVA, as described in the previous algorithm and return the estimated effect size $d^{b}$ and its corresponding p-value $p^{b}$.
  \State Use $d = \frac{1}{B}\sum_{b=1}^{B} d^{b}$ as the estimator and use FDR tool (e.g., \verb+fdrtool+, \verb+ashr+ or \verb+MixTwice+) to adjust those p-values to get the false discovery rate of the estimated effect size.
  \end{algorithmic}
\end{algorithm}

The final step needs large-scale hypothesis testing tools (e.g.,  \citep{zheng2021mixtwice}). Compared to traditional Propensity Score Matching (PSM), bootstrap matching offers greater robustness by employing random subsampling multiple times, which helps to reduce bias and variance in the estimates. Additionally, because the bootstrapping process can be performed in parallel, bootstrap matching can achieve better computational efficiency. This is particularly beneficial since the matching step, after calculating all propensity scores, is computationally intensive. A detailed discussion on this method can be found in \citep{zheng2024bootstrap}.

\subsubsection{causalimpact: under assumption $C_3, C_4$}

\cite{brodersen2015inferring} develops method \verb+causalimpact+ using time series data. The key idea here is to use structural bayesian time series model to predict the treatment-free behavior of experiment group, using the information of control. The gap between the predicted performance and the actual performance of treatment group after the experiment, is therefore the treatment effect. Compared to traditional analysis of covariance (ANCOVA), \verb+causalimpact+ relaxed the assumption of time independence but borrowed more information from longitudinal prospective.

In the \verb+causalimpact+, we always denote $y_t$ be the treatment series and $x_t$ be the control series. We use the following time series model to capture the relationship between control and treatment group:

\begin{equation}
    y_t = \mu_t + \tau_t + \beta'x_t + \epsilon_t
\end{equation}

where $\mu_t$ is the local linear trend that could be modeled using different types of time series model (for example, auto-regressive model, ARIMA model with certain degree of lag, etc), $\tau_t$ is a fixed effect to model seasonality, $\epsilon_t$ is the white noise with certain variance. It is remarkable that some preliminary analysis, such as matching step, is still beneficial but not necessary since we allow the coefficient $\beta \neq 0$ which evaluates the difference between treatment group and control, in the absence of experiment.

\subsection{SPORTSCausal: under assumption $C_4$, $C_5$}

We now try to discuss why \verb+causalimpact+ might not work in the absence of assumption $C_3$.

\begin{figure}[H] % picture
    \centering
    \includegraphics[width=1\columnwidth]{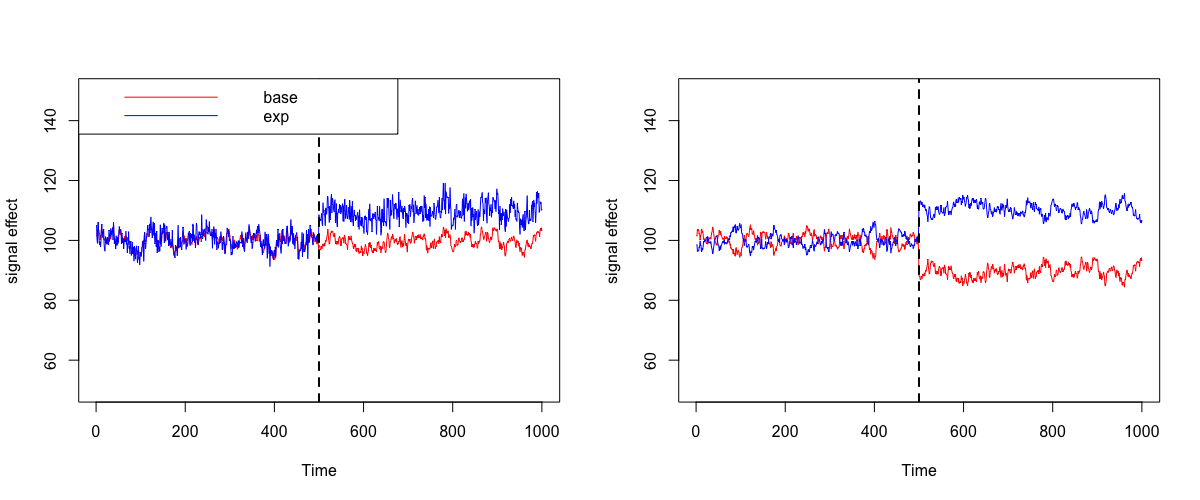}
    \caption{\textbf{The difference when assumption $C_3$ is violated?}}
    \label{demo}
\end{figure}

Figure~\ref{demo} is only a toy example illustrating why the assumption $C_3$ is important for \verb+causalimpact+. The time series is generated using simple auto-regressive model. Left panel of Figure~\ref{demo} shows in the presence of assumption $C_3$. Under the statistical model of \verb+causalimpact+, people uses information from red line (before and after the treatment) to predict the performance of treatment group, without the treatment. Therefore the gap between the predicted mean and the actual mean in after the treatment is the unbiased and consistent estimator of treatment effect. However, in the absence of $C_3$ (right panel), we could suffer from using control group to predict the treatment since there is structural interference between groups when treatment affects. Moreover, the gap between red and blue is the summation of treatment effect and indirect interference effect.

However, when we relax assumption $C_3$ to $C_5$, with the help of $C_4$, we could still get benefits.

\textbf{Claim:}

Under the assumption of $C_4$, $C_5$, treatment effect is identifiable.

The prove of this claim indeed results in the algorithm:

\begin{enumerate}
    \item By assumption $C_4$, we could mimic the distribution of control group, without the presence of treatment interference. This could be fitted using model selection, from a set of time series candidate models. Note that we can still allow some seasnoality effect, besides the local trend, modeling by the time series model.
    \item Assumption $C_5$ is critical for causal identifiability, though it might not influence the algorithm. To be clear, without the assumption $C_5$, the gap between red and blue is not the summation of treatment effect and the treatment interference, but a summation of millions of terms, such as the interference of treatment to control, the interference of control to treatment, the second-round interference of treatment to control, the second-round interference of control to treatment, the third-round interference of treatment to control, etc.
\end{enumerate}

\begin{algorithm}[H]
\label{SPORTSCausal}
  \caption{SPORTSCausal: under assumption $C_4,C_5$}
  \hspace*{0.02in} 
  {\bf Input:}  Outcome matrix $Y$, feature matrix $X$, treatment indicating vector $D$\\
  {\bf Output:} Effect size of treatment $d$ and its significance level $p$
  \begin{algorithmic}[1]
  \State Do Bootstrap Matching outside of the following algorithm:
  \State Fit time series model using $y_{it}, i\leq m, t\leq T_0$, denote as $\mathcal{M}$.
  \State Predict $\hat y_{it}, i\leq m, t\geq T_0 + 1$ using the model $\mathcal{M}$ and replace in the corresponding outcome matrix $Y$.
  \State Run \verb+causalimpact+ using the new data, $X, Y, D$ and report the effect size and corresponding p-value.
  \State Average effect size estimators and use FDR tool (e.g., \verb+fdrtool+, \verb+ashr+ or \verb+MixTwice+) to adjust those p-values to get the false discovery rate of the estimated effect size.
  \end{algorithmic}
\end{algorithm}

\section{Applications}

In online advertising, budget optimization is one of the most critical tasks for maximizing the effectiveness of campaigns. While auction-based systems handle the majority of ad placements, subsidies can sometimes be strategically employed to enhance overall system performance. For instance, one common application of subsidies is the cold start problem, where new advertiser campaigns lack sufficient data for accurately predicting key metrics such as click-through rate (CTR) or conversion rate (CVR). To address this, a subsidy may be assigned to these new campaigns to support model learning, allowing the system to gather the necessary data to improve prediction accuracy over time. By temporarily boosting the visibility of these campaigns, the subsidy helps mitigate the initial disadvantage faced by new advertisers, facilitating a more efficient and effective advertising ecosystem. We call those experiments advertiser campaign budget optimization experiments. 

Even when experiments are correctly randomized at the level of advertisers or ad campaigns, it is challenging to eliminate the spillover effect entirely. For example, in a scenario where the treatment group receives subsidies while the control group does not, the treated campaigns are more likely to win auctions and secure a greater number of ad impressions. This occurs because, with a relatively stable user base, the total number of impressions is unlikely to increase significantly in a short period. Consequently, campaigns with additional budget or subsidies are better positioned to outbid others and capture a larger share of impressions.

Moreover, the impact of the spillover effect can vary depending on the size of the treatment group. When only 5\% of campaigns receive a subsidy, the remaining 95\% may experience reduced opportunities as the subsidized campaigns outcompete them. In contrast, when 50\% of the campaigns receive additional subsidies, the competitive landscape shifts, with half of the campaigns benefiting from the subsidy while the other half faces increased competition. This variation in the spillover effect based on treatment size highlights the complexities involved in designing and interpreting experiments in online advertising, particularly when financial interventions like subsidies are involved.

\begin{figure}[H] % picture
    \centering
    \includegraphics[width=1\columnwidth]{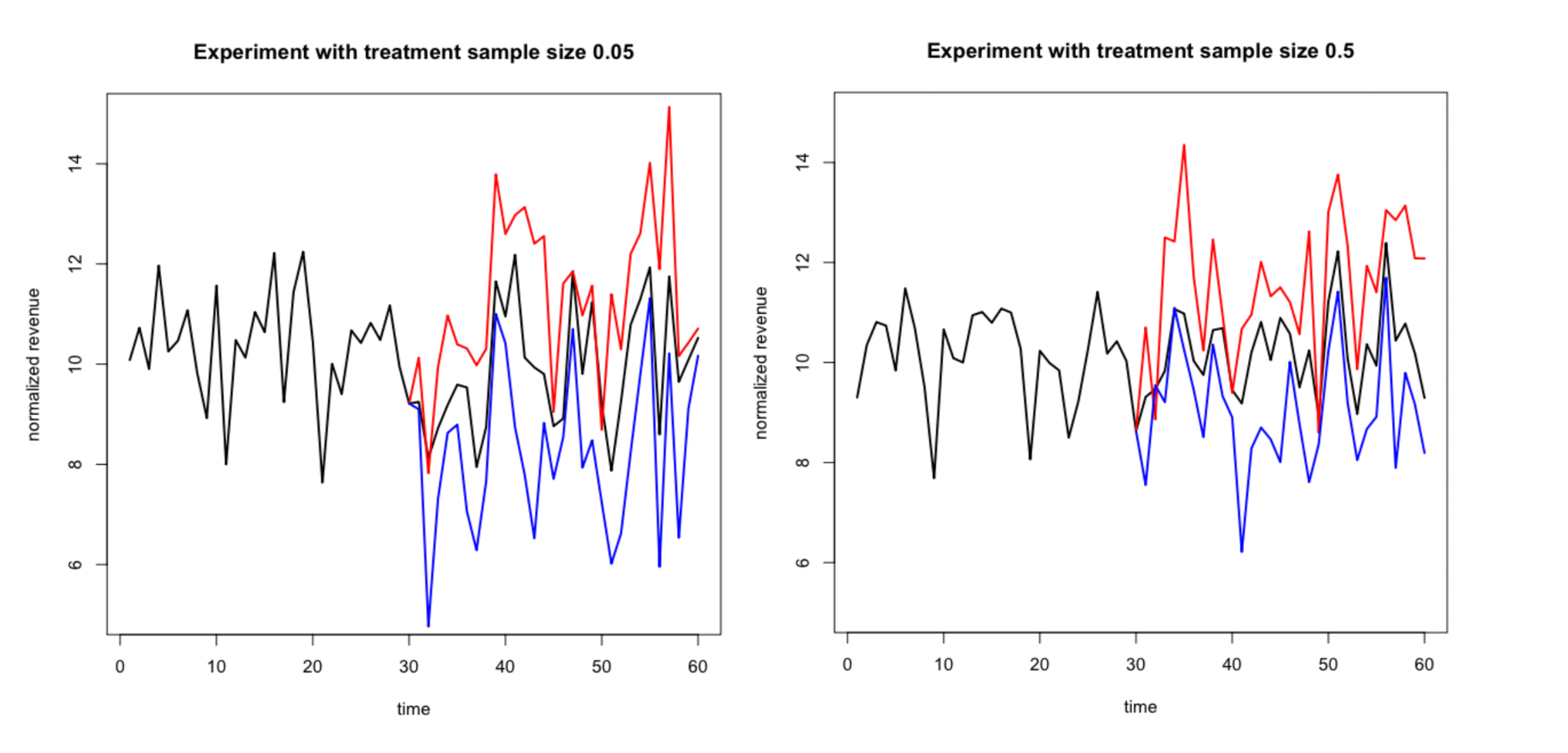}
    \caption{\textbf{Budget optimization experiment under 5\% and 50\% live conditions.} The X-axis represents time, and the Y-axis represents sanitized revenue. Data from the experiment group is shown in red, while the control group is shown in blue. The black curve during the treatment period represents the SPORTSCausal-predicted control.}
    \label{example}
\end{figure}

As shown in Figure~\ref{example}, there is a significant gap between the treatment group and the control group. However, the magnitude of this increase differs notably based on the size of the experiment: the treatment effect is approximately 37.35\% when the experiment size is 5\%, compared to 28.22\% when the experiment size is 50\%. This gap arises not only from the increase in the treatment group but also from a corresponding decrease in the control group. The observed inconsistency in treatment effects between low-traffic and high-traffic experiments, if left uncorrected, can lead to substantial misinterpretations and potentially misguided business decisions. 

Addressing these variations is crucial to ensure that the insights derived from experiments accurately reflect the true impact of interventions across different scales. To achieve this, \verb+SPORTSCausal+ was applied to predict the control (shown in black), resulting in corrected treatment effects of 13.36\% in the 5\% experiment and 14.34\% in the 50\% experiment. By accounting for the spillover effect, \verb+SPORTSCausal+ provides a more consistent and accurate estimation of the treatment effect, ensuring that business decisions are based on reliable data across varying experiment sizes.

The methods applied to more than the settings of online advertising. In the field of economics, particularly in policy evaluation, the assumptions of group independence and randomness are often violated due to interconnected markets and spillover effects. For instance, a government subsidy in one region can influence prices and demand in neighboring regions, complicating the causal interpretation of policy impacts. Traditional econometric models might overestimate or underestimate the true effect of such interventions due to unaccounted spillovers. \verb+SPORTSCausal+ provides a robust framework to account for these complexities, enabling more accurate estimation of policy impacts by adjusting for time-series spillovers. This is particularly valuable in economic studies where interventions can have far-reaching and dynamic effects across different regions or markets. In bioinformatics, high-density peptide arrays, for example, are essential for studying peptide-protein interactions \citep{zheng2020disordered, mergaert2022rheumatoid, amjadi2024novel,parker2024novel}, but their densely packed format can lead to spillover effects, where the binding of one peptide affects neighboring ones. These effects can distort estimates of true binding affinities, complicating the interpretation of results. \verb+SPORTSCausal+ helps address these challenges by accounting for spillover in time-series data, providing more accurate estimates of peptide-antibody interactions. This is crucial for applications like vaccine development, where precise identification of immunogenic peptides is vital. In social and behavioral sciences, particularly in psychology, interventions often influence not only the participants but also those around them, leading to spillover effects. For instance, studies like \citep{liu2022progressive} highlight the importance of retrieval practice in enhancing memory. However, when applying such practices in group settings, spillover effects can occur, where the benefits of one participant's practice may influence others, potentially confounding the results. \verb+SPORTSCausal+ is well-suited to address these complexities by accounting for spillover effects in the analysis of memory interventions. By applying this method, researchers can obtain a clearer picture of the true impact of retrieval practices, ensuring that the observed improvements in memory are accurately attributed to the intervention itself, rather than external influences. This leads to more reliable conclusions that can better inform educational and therapeutic strategies.

\section{Conclusions}

In this paper, we have introduced \verb+SPORTSCausal+ and its R package \citep{sports}, a novel approach to causal inference that addresses the challenges posed by the absence of group randomness and the presence of spillover effects. Traditional methods, such as Propensity Score Matching (PSM) and other matching techniques, often rely on strong assumptions that may not hold in real-world scenarios, leading to biased estimates and misleading conclusions. \verb+SPORTSCausal+ overcomes these limitations by accounting for both direct and indirect effects, providing a more accurate assessment of treatment impacts.

We demonstrated the effectiveness of \verb+SPORTSCausal+ through a real-world budget-control experiment, where traditional methods failed to capture the true treatment effect due to significant spillover. By correcting for these spillover effects, \verb+SPORTSCausal+ delivered robust and reliable estimates across different experiment scales. Our discussion extended beyond business applications, highlighting the broader relevance of \verb+SPORTSCausal+ in fields such as economics, bioinformatics, and social and behavioral sciences. 

Overall, \verb+SPORTSCausal+ represents a significant advancement in causal inference, offering a flexible and robust tool for researchers across various domains. However, one limitation of the proposed approach is its lack of integration with experimental design principles. While \verb+SPORTSCausal+ provides substantial benefits in real-world settings where experiment setups are less than ideal, we believe that combining this method with efficient experimental design could further enhance its effectiveness.

In the context of online advertising, for example, one of the critical challenges is determining the minimum traffic needed to accurately estimate the true treatment effect within a reasonable timeframe. Many trials are exploratory, and diverting significant live traffic—especially when it's user-visible—is often undesirable. Simultaneously, business decisions must be made swiftly, meaning teams cannot afford to wait for data to fully converge. Given these constraints, along with the complexities introduced by spillover effects, we see significant opportunities for future research to refine and extend \verb+SPORTSCausal+, particularly in integrating it with optimized experimental design strategies.

\clearpage
\bibliography{references}

\end{document}